\documentclass[aps,pra,amsmath,amssymb,amsfonts,superscriptaddress,lengthcheck,twocolumn,longbibliography]{revtex4-1}
\usepackage{dcolumn}    
\usepackage{graphicx}
\usepackage{amsmath, bbm}
\usepackage{latexsym}
\usepackage{amsfonts}   
\usepackage{amssymb}
\usepackage{array}      
\usepackage{epsfig}
\usepackage{txfonts}
\usepackage{xcolor,braket}
\usepackage[colorlinks=true,linkcolor=blue,urlcolor=blue,citecolor=blue,pdfusetitle]{hyperref}
\usepackage{hyperref}
\usepackage{ulem}
\usepackage{soul}
\usepackage{dsfont}

\usepackage{relsize}

\usepackage{mathrsfs}

\usepackage{cancel}
\usepackage{enumitem}

\providecommand{\bgreek}[1]{\mbox{\boldmath$#1$}}

\newcommand{\dyad}[1]{\left\vert#1\right\rangle\!\left\langle#1\right\vert}
\newcommand{\tr}[1]{\mathrm{tr}\left\{#1\right\}}

\begin{document}
\title{Action quantum speed limits}

\author{Eoin O'Connor}
\affiliation{School of Physics, University College Dublin, Belfield Dublin 4, Ireland}

\author{Giacomo Guarnieri}
\affiliation{Department of Physics, Trinity College Dublin, Dublin 2, Ireland}
\affiliation{Dahlem Center for Complex Quantum Systems,
Freie Universit\"{a}t Berlin, 14195 Berlin, Germany}
       
\author{Steve Campbell}
\affiliation{School of Physics, University College Dublin, Belfield Dublin 4, Ireland}

\begin{abstract}
We introduce action quantum speed limits (QSLs) as a family of bounds on the minimal time to connect two states that, unlike the usual geometric approach, crucially depend on how the path is traversed, i.e. on the instantaneous speed. The two approaches provide consistent bounds when the instantaneous speed is optimized along a fixed path and we demonstrate this explicitly for the case of a thermalizing qubit employing techniques from optimal control theory. In addition, we critically analyze the interpretation of QSLs based on different choices of metric establishing that, in general, these open system QSL times provide an indication of optimality with respect to the geodesic path, rather than necessarily being indicative of an achievable minimal time.
\end{abstract}
\date{\today}

\maketitle

\section{Introduction}
\label{Sec:Intro}
Time has always proved to be a difficult physical concept to grasp. Its intrinsic directionality, commonly quantified through the notion of entropy production, allows to uniquely define a `before` and an `after`, and therefore time dictates the evolution of a system, i.e. its transformation between different configurations (or states). The theory of quantum computation, for one, considers time a key resource to optimize, since it relates to the number of elementary computational operations that can be performed~\cite{LloydNature}. A basic question naturally arises: Is there a lower bound on the time it can take to transform between a given initial and final target states? The ubiquity of this simple inquiry has stimulated significant research efforts, aimed at determining the ultimate limits on a physical evolution.

The laws of quantum mechanics allow to introduce a fundamental bound on the speed of evolution. Among the first to realize this were Mandelstam and Tamm~\cite{Mandelstam1945} (MT), who in their seminal work showed that the minimum amount of time for a quantum system to unitarily evolve from a pure state $\ket{\psi_i}$ to an orthogonal state $\ket{\psi_f}$ is given by $\tau \geq \tau^{MT}_{QSL}\!=\!(\pi/2) \frac{\hbar}{\Delta H}$, where $(\Delta H)^2$ is the variance of the energy of the system in the initial state. This revolutionary re-interpretation of the energy-time uncertainty relation was termed the quantum speed limit (QSL) time. Their result was later generalized to arbitrary initial states by Bhattacharyya~\cite{Bhattacharyya1983, Fleming1973}, who proved that
\begin{equation}
\label{QSLMT}
\tau\geq \tau_\text{QSL} = \frac{\hbar \arccos(|\left\langle \psi(0)|\psi(\tau)\right\rangle|)}{\Delta H}. \end{equation}
Subsequently, Anandan and Aharonov~\cite{Anandan1990} further extended the situation to time-dependent Hamiltonians and, most importantly, showed that Eq.~\eqref{QSLMT} could be understood from a purely geometric perspective as a consequence of the properties of the Fubini-Study metric on the Riemannian manifold of quantum states. In particular, they showed that in this metric, the geodesic length is given by the Bures angle and the path length of any unitary dynamics is given by $\int_0^\tau dt \Delta H(t)/\hbar$. Therefore, the MT bound simply follows from the fact that the path length cannot be smaller than the geodesic length. An alternative QSL bound, valid under the same assumptions, was derived by Margolus and Levitin~\cite{Margolus1998} (ML), although crucially involving the mean energy $\langle H \rangle$ of the initial state instead of the variance $\Delta H$: $\tau \geq \tau^{ML}_{QSL}\!=\!\pi \hbar/ 2 \langle H \rangle$. Importantly, Levitin and Toffoli~\cite{Levitin2009} later proved that both bounds $\tau^{MT}_{QSL}, \tau^{ML}_{QSL}$ are only attainable when $\Delta H \!=\! \langle H \rangle$ and therefore become equivalent when saturated. Understanding and developing these results has been the focus of sustained work~\cite{Wootters1981,Braunstein1994,Brody2006,Deffner2013Unitary,ModiPRL,Frowis2012,Poggi2013,PoggiPRA,Giovannetti2004,Mondal2016PLA,Campbell2017,Marvian2015,Marvian2016, AdolfoNJP, AdolfoPRL2018, Mostafazadeh2009,FogartyPRL2020,Lychkovskiy1,RicardoPRR2020,PoggiArXiv,FelixPRA2020}.

Extending the notion of QSLs to open quantum systems has, however, proven to be a more formidable task, with the geometric approach bearing the most fruit~\cite{Pires2016,Taddei2013,DelCampo2013,Deffner2013Open,SunSciRep,Mondal2016,Jones2010,Zhang2014,Mirkin2016,Uzdin2016,Luo2018,Maniscalco2019,Ektesabi2017,Cai2017,Deffner2017,WuYuPRA,Funo2019,Brody2019,Campaioli2019,VanVu2020,Campaioli2020,DazV2020}, see also the reviews~\cite{Frey2016, DeffnerReview}. Arguably the main hurdle in moving to the open system setting is that there is an infinite family of suitable metrics for mixed states describing open quantum systems to choose as a distance measure, as characterized by the Morozova, {\v C}encov, and Petz (MCP) theorem~\cite{Morozova1991, Petz1996}. This is in contrast to pure states undergoing unitary dynamics where the Fisher information metric is the sole contractive Riemannian metric that can be defined in order to measure the distance between states, thus enforcing Eq.~\eqref{QSLMT} as the only saturable geometric bound under these conditions. However, for open systems any given choice of metric leads to a bona-fide QSL. While Pires~\textit{et al.}~\cite{Pires2016} derived an infinite family of geometric QSLs, they also crucially showed that, for any fixed path connecting two arbitrary (initial and final) states, it was possible (at least in principle) to identify the metric that gives rise to the tightest QSL bound. However, fixing the path and the end points in the geometric approach completely determines all the quantities and therefore, unless the given path already coincides with a geodesic according to some metric tensor, the theoretical lower bound for the QSL time is not achievable.

In this work, we highlight that the speed at which a path is traversed represents a further meaningful parameter for optimization of the dynamics. Curiously, geometric QSLs are insensitive to this degree of freedom, since by construction they only rely on the time averaged speed. Therefore, we introduce a different approach to the problem -- the \textit{action} QSL -- showing that their saturability depends both on the path taken and on the speed at which that path is traversed. Geometric QSLs are shown to coincide with special instances of the action QSLs, thus demonstrating the consistency of our result with that of Ref.~\cite{Pires2016}, however, while the geometric formulation will imply that all ramp profiles along a path are equivalent, the action QSL is sensitive to the instantaneous speed. Due to their construction, finding the best way to traverse a path is naturally suited to be tackled with quantum optimal control theory~\cite{glaser2015}. This is explicitly shown in a paradigmatic example of a qubit thermalizing with an environment modelled using a generalized amplitude-damping channel. The implementation of standard optimal control techniques allows us to find the optimal way to travel along the path.

The paper is organized as follows. Section~\ref{sec:geomQSL} is devoted to a brief review of geometric QSLs, highlighting some main conceptual limitations which are then elucidated through a simple example of a qubit subject to a generalized amplitude damping channel. In Section~\ref{sec:ASL} we introduce action quantum speed limits as an alternative to the geometric approach and establish that they provide consistent bounds when the path is optimally traversed. In Section~\ref{sec:OC} we explicitly illustrate this by combining action QSLs with optimal control techniques. Finally, our conclusions and some further discussions are are presented in Section~\ref{sec:Concl}.

\section{Geometric Quantum Speed Limits}
\label{sec:geomQSL}
The geometric approach to deriving quantum speed limits in essence relies on the simple and elegant consideration that the geodesic distance between any two points of a Riemannian metric is the shortest possible length connecting them. In particular, in the case of open quantum systems, such a manifold, which we denote by $\mathcal{D}$, is represented by the set of statistical operators, $\rho$, over a given Hilbert space, $\mathscr{H}$, i.e. the convex set of positive semi-definite and trace-1 linear operators acting on $\mathscr{H}$. The MCP theorem~\cite{Morozova1991, Petz1996} characterizes the whole family of Riemannian metrics $\{g\}$ on $\mathcal{D}$ that are contractive under the action of physical channels, i.e. completely positive and trace-preserving (CPTP) maps, and can therefore be used to distinguish between two states. Since in the framework of QSLs we are interested in finding the minimum time required to evolve an initial state into a final target state, let us parametrize the density matrices by $\rho({\bgreek{\lambda}})$, where $ \bgreek{\lambda}\in\mathbb{R}^M$ denotes a set of time-dependent control parameters which are changed according to some given smooth protocol. Upon choosing the parametrization $t\in[0,\tau] \to \bgreek{\lambda}(t)$, with $\tau$ denoting the total evolution time, a protocol which changes smoothly $\bgreek{\lambda}_I \equiv \bgreek{\lambda}(0)$ to $\bgreek{\lambda}_F \equiv \bgreek{\lambda}(\tau)$ geometrically draws a path $\gamma$ onto the Riemannian manifold of quantum states that connects $\rho_0 \equiv \rho(\bgreek{\lambda}_I)$ and $\rho_\tau \equiv \rho(\bgreek{\lambda}_F)$.

The length of the path is then obtained by integrating the metric-induced infinitesimal length along the curve $\gamma$, i.e.
\begin{equation}\label{PathLength}
\ell_{g}^{\gamma}(\rho_0,\rho_\tau) = \int_{\gamma} ds = \int_{0}^{\tau} dt\,\sqrt{\sum_{jk=1}^M g_{jk} \frac{d\lambda_j}{dt}\frac{d\lambda_k}{dt}}.
\end{equation}
Being the shortest path, the length of the corresponding geodesic curve is defined as 
\begin{equation}\label{defgeodesic}
\mathcal{L}_{g}(\rho_0,\rho_\tau)  = \min_{\gamma} \ell_{g}^{\gamma}(\rho_0,\rho_\tau) .
\end{equation}
The family of geometric QSL stems precisely from Eq.~\eqref{defgeodesic}, restated as
\begin{equation}\label{path_len}
\mathcal{L}_{g}(\rho_0,\rho_\tau)  \leq \ell_{g}^{\gamma}(\rho_0,\rho_\tau).
\end{equation}
It is important to stress that Eq.~\eqref{defgeodesic} expresses a hierarchy among all possible paths connecting the two states $\rho_0$ and $\rho_\tau$ for a \textit{fixed} metric $g$. 

In order to translate this inequality into a QSL for the evolution time as in Eq.~\eqref{QSLMT}, one usually introduces the path-average speed
\begin{equation}
\label{gen_speed}
v_g^\gamma =\frac 1\tau \ell_{g}^{\gamma}(\rho(0),\rho(\tau)),
\end{equation}
from which it straightforwardly follows that
\begin{equation}
\label{GEOMETRICQSL}
\tau \geq \tau_g^\gamma =\frac{\mathcal{L}_{g}(\rho(0),\rho(\tau))}{v_g^\gamma}.
\end{equation}
Saturating Eq.~\eqref{path_len} is equivalent to $\tau \!=\! \tau_g^\gamma$ and this occurs when the dynamics follows the corresponding geodesic path for a given metric $g$. As already mentioned, for pure states and unitary evolution, the Fisher information metric represents the unique contractive Riemannian metric $g$ and leads to the MT bound. Furthermore, in this case the speed is related to a physical resource of the system, namely the (square root of the) energy variance of the initial state. Whenever open quantum systems and mixed states are considered, however, the non-uniqueness of $g$ naturally brings forward an important question: is there a particular metric which gives rise to a QSL which is the tightest possible, therefore representing the ultimate lower bound on the evolution time?

The answer to the above question is actually very subtle. In Ref.~\cite{Pires2016} it was argued that, for any given path $\gamma^*$ between two fixed initial and final states $\rho_0,\rho_{\tau}$, the hierarchy of the MCP metrics reflects into the possibility to find, at least in principle, the geodesic which gives rise to the tightest geometric QSL bound to the evolution. The latter is given by
\begin{equation}\label{minGeomQSL}
\tau_{\mathrm{QSL}} = \tau_{g^*}^{\gamma^*}  \leq \tau,
\end{equation}
where the metric $g^*$ is the one such that its geodesic $\mathcal{L}_{g^*}(\rho(0),\rho(\tau))$ is the closest to the actual given path $\gamma^*$, i.e.
\begin{equation}\label{minGeomQSLCONDITION}
    g^*\,\text{such that}\, \inf_g \delta_g^{\gamma^*} = \delta_{g^*}^{\gamma^*}, 
\end{equation}
with $\delta_g^{\gamma^*} \equiv \tau/\tau_g^{\gamma^*}-1$.

%\textbf{Q: Can we formalize it in an Equation similar to the one above but for the time [that would be best, since we focus on that] or do we have to introduce necessarily here the concept of resource/constraint?\color{blue}above would be the $\tau$ formulation}

This result highlights that geometric QSLs should be carefully interpreted. Once all the quantities entering the bound Eq.~\eqref{minGeomQSL} are uniquely determined, i.e. once a path $\gamma^*$ and start and end points are fixed, nothing more can be done in order to approach the QSL bound. Equivalently said, if a given path connecting two quantum states is not already optimal, in the sense that does not already coincide with a geodesic path according to some contractive Riemannian metric, then the geometric QSL bound is never saturable and will only provide an estimate of ``how  far from optimal" the evolution time is with respect to $\tau_{\mathrm{QSL}}$. 

Thus, if we are free to choose the path connecting a given initial and target state, then the QSL bound for every metric $g$ can be saturated simply by moving along a path which coincides to the geodesic for that metric. {\it A priori}, the choice of one metric over another may be dictated by the physics of the problem at hand, e.g. the average initial energy as in the ML bound or the initial energy variance as for the MT bound. Regardless though, the corresponding QSL bound can, in principle, be achieved. Conversely, however, when a particular dynamics is considered, i.e. one dictated by a specified CPTP map, then there is no single metric $g$ which represents the tightest QSL for every possible choice of the path's boundary conditions.

Let us provide an explicit demonstration of the above for a simple paradigmatic example consisting of a qubit undergoing a generalized amplitude damping channel (GADC). This rather ubiquitous situation describes a two level quantum system undergoing equilibration with a large thermal bath, such that its evolution is described in terms of the following master equation (in interaction picture)
\begin{multline}
\label{thermME}
\dot{\varrho}_{S} = \mathcal{L}(\varrho_{S}) = \gamma \left( \sigma_- \varrho_{S}  \sigma_+ - \frac{1}{2} \Big\{ \varrho_{S} , \sigma_+ \sigma_-  \Big\} \right) \\+ \Gamma \left( \sigma_+ \varrho_{S} \sigma_- - \frac{1}{2} \Big\{ \varrho_{S} , \sigma_- \sigma_+ \Big\} \right)
\end{multline}
where $\beta\!=\tfrac{1}{2}\!\ln \tfrac{\gamma}{\Gamma}$ denotes the inverse temperature of the bath in units of the qubit's energy.

\begin{figure}[t]
\hskip0.1\columnwidth(a) \hskip0.4\columnwidth(b)
\includegraphics[width=0.58\columnwidth]{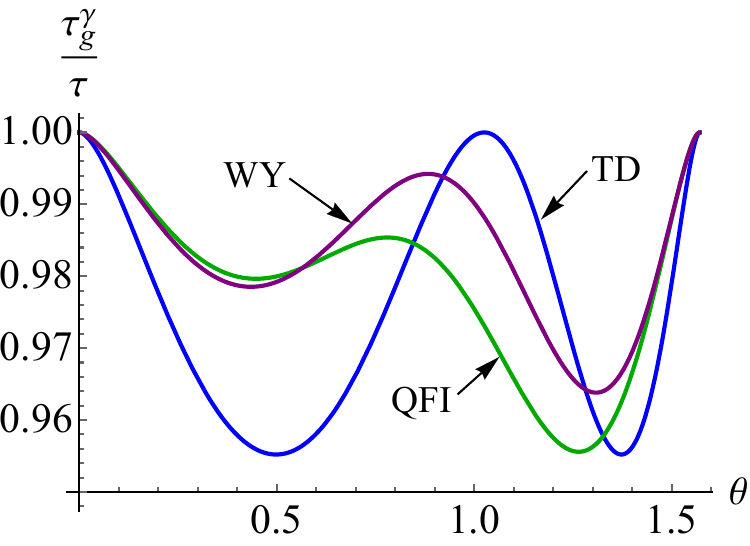}~~\includegraphics[width=0.38\columnwidth]{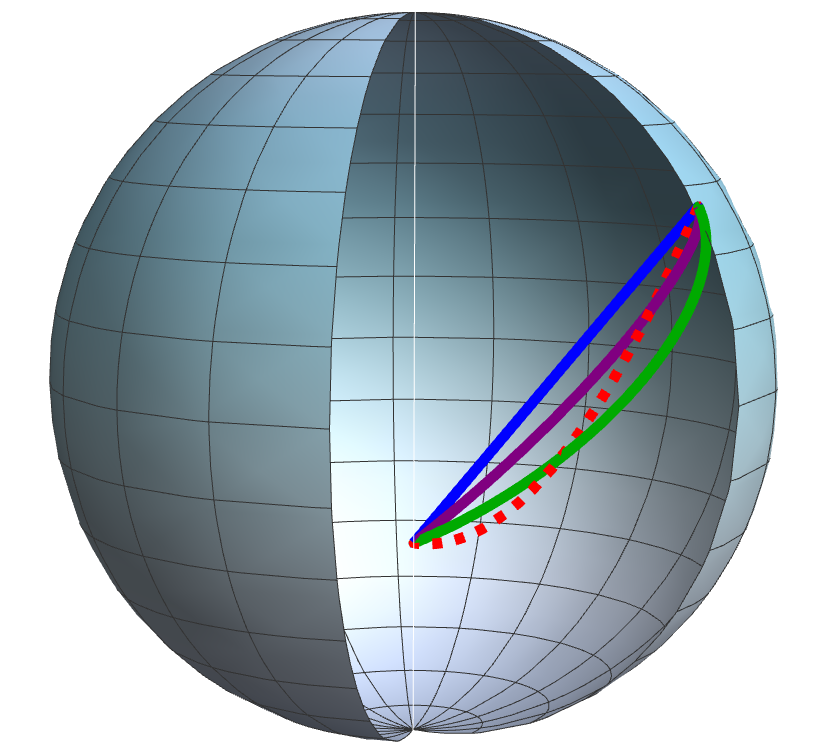}
\caption{
\textbf{(a)} Plot of $\tau^\gamma_g/\tau$, with $\beta\!=\!0.5$, for three different choices of metric $g = \mathrm{WY},\mathrm{QFI},\mathrm{TD}$ (Wigner-Yanase, quantum Fisher information and trace distance, respectively), as a function of the initial state parameter $\theta$. The path $\gamma$ corresponds to the GADC. \textbf{(b)} Bloch sphere representation of the various paths for a two-level system connecting $\ket{\Psi_0}\!=\! \frac{1}{2}\ket{0}+\tfrac{\sqrt{3}}{2}\ket{1}$ and the steady state of Eq.~\eqref{thermME} with $\beta\!=\!0.5$. The innermost straight, blue line corresponds to the TD geodesic path. The QFI geodesic corresponds to the outermost solid, green curve. The remaining solid, purple line is the WY geodesic. Finally, the path followed by the GADC is shown in dashed, red.}
\label{fig:Paths}
\end{figure}

While we will consider the dynamics (GADC) and the final state (i.e. the thermal state) fixed, we will vary the initial state $\rho_0$, thus resulting in a different path on the Bloch sphere for each starting configuration. We will focus on three important metrics for which their geodesics can be calculated and compute the respective QSL bound Eq.~\eqref{GEOMETRICQSL}.
The first is based on the quantum Fisher information (QFI)~\cite{Wootters1981,Braunstein1994,Paris2009}. The geodesic distance in this metric is given by the Bures angle
\begin{equation}
\mathcal{L}_\text{QFI}(\rho,\sigma)=\arccos(\text{tr}\sqrt{\sqrt{\rho}\sigma\sqrt{\rho}}).
\end{equation} 
The quantum Fisher information is defined as the expectation value of the square of the symmetric logarithmic derivative operator $L$, $\mathcal{F}_Q\!=\!\text{tr}[\rho L^2] $ where $L$ is defined implicitly by $\dot{\rho}\!=\!(\rho L +L \rho)/2 $. The resulting QSL reads~\cite{Taddei2013}
\begin{equation}
    \tau_{\text{QFI}} = \frac{\mathcal{L}_\text{QFI}}{\frac{1}{\tau}\int_0^\tau dt \sqrt{\mathcal{F}_Q}}.
\end{equation}
The second metric we will consider is the one based on the Wigner-Yanase (WY) skew information metric~\cite{Pires2016}. The geodesic distance in this metric is given by
\begin{equation}
\mathcal{L}_\text{WY}(\rho,\sigma)=\arccos(\text{tr}(\sqrt{\rho}\sqrt{\sigma})).
\end{equation}
This geodesic distance was derived by Gibilisco and Isola~\cite{GIBILISCO2001,Gibilisco2003} and is a quantum generalisation of the Bhattacharya angle.
This metric is known as the WY metric because for unitary dynamics with time dependent Hamiltonian $H_t$ we have~\cite{Pires2016,Gibilisco2003}
\begin{equation}
    \ell_{WY}^\gamma(\rho_0,\rho_\tau) = \sqrt{2}\frac{1}{\tau}\int_0^\tau dt\sqrt{\mathcal{I}(\rho_t,H_t)}
\end{equation}
where $\mathcal{I}(\rho,A)\!=\!-(1/2)\tr{[\!\sqrt{\rho},A]^2}$ is the WY skew information of the self-adjoint matrix $A$. 
%\textbf{This is a kind of condensed version of the reasoning given by Gibilisco~\cite{Gibilisco2003} on page 3758 but it's probably the best I can do without adding another appendix section}. 
Finally, we will consider the metric based on the trace distance (TD)~\cite{Cai2017} which stems from a direct application of the triangle inequality~\cite{Deffner2017}. The distance is given by the trace norm
\begin{equation}
\mathcal{L}_\text{TD}(\rho,\sigma)=\| {\rho(0)-\rho(\tau)} \|_1 = \text{tr}\sqrt{(\rho(0)-\rho(\tau))^2}.
\end{equation}
The geodesic paths are also known for all of these three metrics, for example, the TD geodesic path is simply a ``straight line" between the initial and final states
\begin{equation}
\rho(t)=(1-p(t))\rho(0)+p(t)\rho(\tau),
\label{TRgeoeq}
\end{equation}
where $p(t)$ is any function satisfying $p(0) \!=\! 0$ and $p(\tau) \!=\! 1$ that is monotonically increasing on that interval. The paths for the QFI and WY metrics were derived by Uhlmann~\cite{Uhlmann1995} and Gibilisco~\cite{Gibilisco2003}, respectively, and are explicitly provided in the Appendix A. Exemplary paths for a qubit are shown in Fig.~\ref{fig:Paths}(b).

%%%%%%%%%%%%%%%%%%%%
%%%%%%%%%%%%%%%%%%%%%

Fig~\ref{fig:Paths}(a) shows the result of the evaluation of the three QSL bounds as a function of the parameter $\theta$ which determines the initial state that, without any loss of generality, is taken to be pure $\rho_0(\theta)\!=\!\ket{\Psi_0(\theta)}\bra{\Psi_0(\theta)}$, with $\ket{\Psi_0(\theta)}\!=\!\cos(\theta)\ket{0} + \sin(\theta)\ket{1}$. While for any given fixed path (i.e. in this case, fixed $\theta$), one QSL is clearly tighter than the other, thus confirming Eq.~\eqref{minGeomQSL}, it is evident that none of them provides the tightest QSL for every parameter $\theta\in [0,\pi]$. The tightest bound for any given initial state is the one corresponding to the metric whose geodesic happens to be closest to the GADC dynamics for that particular choice of $\theta$. We finally notice that for $\theta\!=\! 0, \pi$ all the bounds saturate because the GADC becomes equivalent to the depolarising channel, which traces a geodesic path for all of the considered metrics. 

\section{Action Quantum Speed Limits}
\label{sec:ASL}
The previous section highlights that when open quantum systems are considered, there is no single Riemannian contractive metric for which the corresponding geometric QSL bound is the tightest unless the path and endpoints are fixed. Fixing a path is therefore a necessary requirement in order to have a well-defined unique and tightest QSL bound Eq.~\eqref{minGeomQSL} such that Eq.~\eqref{minGeomQSLCONDITION} holds. If every parameter of the problem is fixed, however, unless a given dynamics already coincides with a geodesic path according to some metric (e.g. in the case of a depolarizing channel, see $\theta=0,\pi$ in the above example), then the geometric QSL time provides a quantitative indication of how far the traversed path is from the optimal path, according to the specific metric in question. Nevertheless, in spite of what the name might suggest, the geometric quantum speed limit time is completely insensitive to the actual instantaneous speed. 

This simple observation represents the starting point for introducing our new family of QSLs. The instantaneous speed at which a given path is travelled is an important degree of freedom. Indeed, the varying speed of evolution provides a vital tool in many physical settings, for example every thermodynamic cycle of any driven engine is crucially dependent on the speed at which the protocol is performed~\cite{MarkPRX} and high-fidelity control can be achieved by varying the speed with which some time-dependent ramp is applied such that the dynamics slows down when energy gaps close which applies to understanding the dynamics across quantum phase transitions in light of the Kibble-Zurek mechanism~\cite{RicardoPRR2020}.

We incorporate the instantaneous speed into the formulation of quantum speed limits by borrowing inspiration from recent developments in thermodynamic geometry~\cite{SalamonPRL,Crooks2007,Zulkowski2012,Feng2009,Sivak2012,Scandi2019,MarkPRX,Scandi2020}. This can be achieved by applying the Cauchy-Schwarz inequality $\int_0^\tau h^2dt\int_0^\tau f^2dt\geq[\int_0^\tau f h \,dt]^2$ to the path length in equation~\eqref{path_len}. Specifically, by setting $h\! =\! 1$, one has
\begin{equation}
    \tau \int_0^\tau dt\,\sum_{jk=1}^M g_{jk} \frac{d\lambda_j}{dt}\frac{d\lambda_k}{dt} \geq \left(\int_0^\tau dt \sqrt{\,\sum_{jk=1}^M g_{jk} \frac{d\lambda_j}{dt}\frac{d\lambda_k}{dt}}\right)^2
\end{equation}
which leads to the following result
%by setting $h\! =\! 1$ and $f\! =\! ds$ (with $ds$ being defined in Eq.~\eqref{PathLength}), we obtain $\tau\int_0^\tau ds^2 \geq \left(\int_0^\tau ds\right)^2$ which leads to the following
\begin{equation}\label{ACTIONQSL}
    \tau\geq\tau_{a}^{\gamma} = \frac{\mathcal{L}_{g}(\rho(0),\rho(\tau))^2}{a_g^\gamma},
\end{equation}
where $a_{g}^{\gamma} \!=\! \int_0^\tau dt\,\sum_{jk=1}^M g_{jk} \frac{d\lambda_j}{dt}\frac{d\lambda_k}{dt}$ possesses the dimensions of an action. Eq.~\eqref{ACTIONQSL} is the anticipated new family of QSLs which, in light of the above quantity and its interpretation, we name \textit{action quantum speed limits}. We remark that action QSLs do not suffer some of the issues associated with geometric QSLs. In particular, as highlighted in Ref.~\cite{Mirkin2016}, for the damped Jaynes-Cummings model, geometric QSLs may diverge as the total process time increases, despite the fact that the distance between initial and steady state is fixed and, furthermore, the time required to approach infinitesimally close to the steady state being finite. In contrast, due to their construction, action QSLs remain bounded and therefore capture the essence of the original QSL formulations more aptly~\cite{Mirkin2016}.

It is crucial to point out that, for any given path $\gamma^*$, Eq.~\eqref{ACTIONQSL} is saturated when $\gamma^*$ is a geodesic and the speed along it, $\sqrt{\,\sum_{jk=1}^M g_{jk} \frac{d\lambda_j}{dt}\frac{d\lambda_k}{dt}}$ is constant. This means that, if a given path is already optimal in the sense that it coincides with a geodesic and thus saturates the geometric QSL, then the action QSL will also be saturated provided this path is traversed at a constant speed in the corresponding metric. Conversely, when a given path is not optimal, then Eq.~\eqref{ACTIONQSL} becomes more delicately dependent on this instantaneous speed, as we show explicitly below.

Another important consideration stems from the following chain of inequalities:
\begin{equation}
    \frac{\tau_{a}^{\gamma}}{\tau} = \frac{\mathcal{L}_{g}(\rho(0),\rho(\tau))^2}{\tau \,a_g^\gamma}\leq \frac{\mathcal{L}_{g}(\rho(0),\rho(\tau))^2}{ \,\tau^2(v_g^\gamma)^2} = \left(\frac{\tau_{g}^{\gamma}}{\tau}\right)^2
    \label{CSineq}
\end{equation}
This result highlights the fact that the geometric QSL is a special instance, corresponding to the upper bound, of the action QSL. This physically indicates that, for every non-optimal path, any time-dependent profile for the speed will lead to a QSL bound which is going to be less than or equal to the geometric ideal QSL value. This is however a very important property, as it implies that different strategies aimed at optimizing the speed for any given non-optimal path will reflect in the value of the action-QSL time which progressively approaches the bound given in Eq.~\eqref{CSineq}. Due to the very structure of it involving the action $a_{g}^{\gamma}$, finding the optimal way to traverse a path is naturally suited to be solved by techniques borrowed from optimal control theory~\cite{Deffner2014,MukherjeePRA,Zulkowski2015,Deffner2020,Miller2019,glaser2015,liu2017quantum,mirkin2020quantum,basilewitsch2020optimally}.

\subsection*{Optimizing the Instantaneous Speed}
\label{sec:OC}
In order to identify the optimal time-dependent profile for the dynamics we can make use of Pontryagin's optimum principle~\cite{Kirk2004, Deffner2014} to find an effective control Hamiltonian that realizes a particular dynamics while minimising a given action. A full explanation of the optimal control procedure can be found in Appendix B. In order to explicitly show this, we will consider the same example as Section~\ref{sec:geomQSL}, a qubit subject to a generalized amplitude damping channel. To explicitly account for the time-dependence following the path we express this channel using the Kraus operators
\begin{multline}
    K_0(t) = \sqrt{c}\begin{pmatrix}\sqrt{1-p(t)}&0\\0&1\end{pmatrix},K_1(t) = \sqrt{c}\begin{pmatrix}0&0\\\sqrt{p(t)}&0\end{pmatrix},\\
    K_2(t) = \sqrt{1-c}\begin{pmatrix}1&0\\0&\sqrt{1-p(t)}\end{pmatrix},K_3(t) = \sqrt{1-c}\begin{pmatrix}0&\sqrt{p(t)}\\0&0\end{pmatrix},
\end{multline}
where $c\!=\!\frac{1}{2}(1+\tanh{\beta})$ and with $\beta$ being the inverse temperature of the bath. The path is fixed by the value of $\beta$, while $p(t)$ describes how that path is traversed. It is important to stress that to optimise the dynamics we must saturate the Cauchy-Schwarz inequality, which corresponds to finding the ramp profile that results in a constant speed in the metric. As we demonstrate by explicit example for non-geodesic paths, achieving a constant speed in the metric generally requires a non-trivial temporal ramp profile.

%For the optimal control our external control parameter is $\dot{p}(t)$ and the condition for an admissible control is $\int_0^\tau dt \dot{p}(t)\!=\!p(\tau)-p(0)$~\cite{Deffner2014}. This ensures that our start and endpoints are fixed.

\begin{figure}[t]
\hskip0.02\columnwidth(a) \hskip0.4\columnwidth(b)\\
\includegraphics[width=0.5\columnwidth]{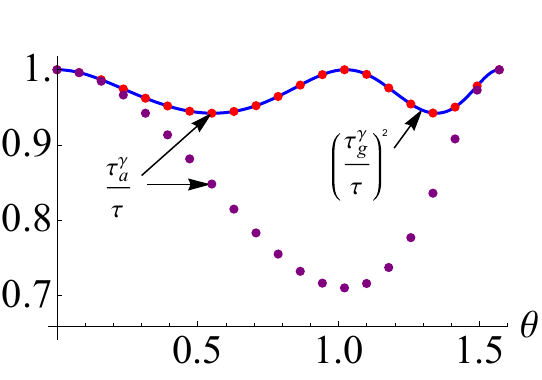}~\includegraphics[width=0.5\columnwidth]{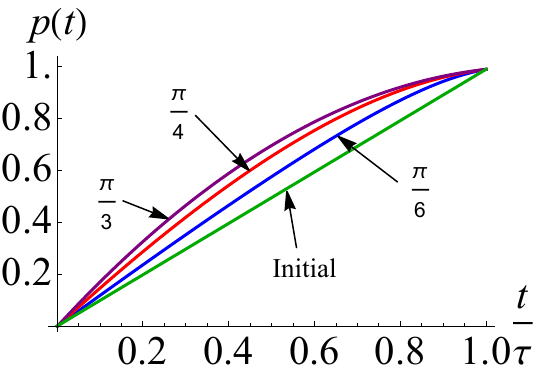}\\
\hskip0.02\columnwidth(c) \hskip0.4\columnwidth(d)\\
\includegraphics[width=0.5\columnwidth]{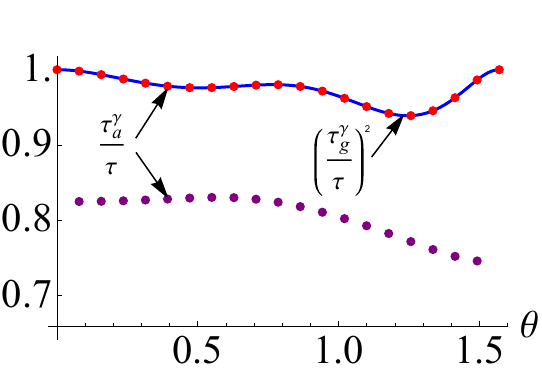}~\includegraphics[width=0.5\columnwidth]{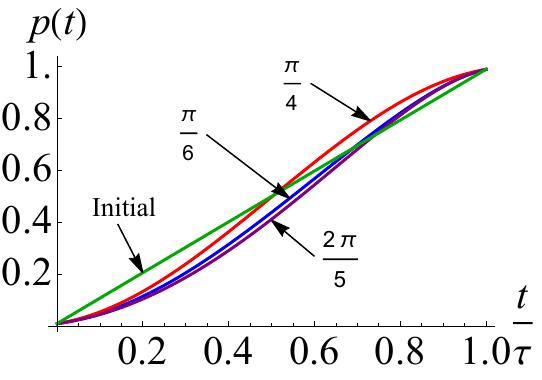}\\
\hskip0.02\columnwidth(e) \hskip0.4\columnwidth(f)\\
\includegraphics[width=0.5\columnwidth]{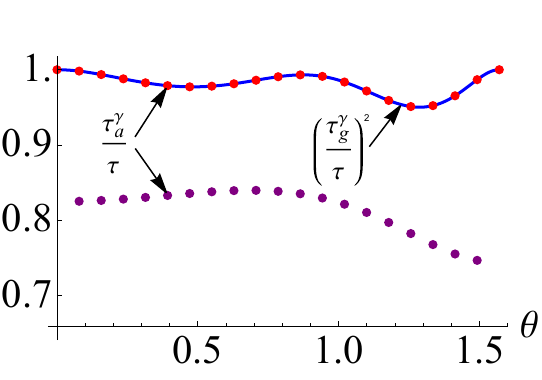}~\includegraphics[width=0.5\columnwidth]{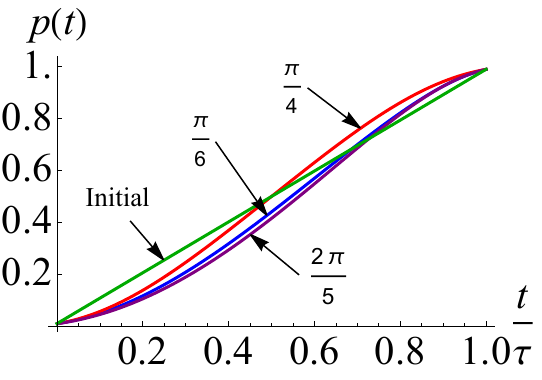}
\caption{
\textbf{(a)} The solid blue line shows the (square of the) geometric TD speed limit for the generalised amplitude damping channel with $\beta = 0.5$ (arbitrarily chosen). This QSL time is independent of $p(t)$ as long as $\dot{p}(t)\!>\!0$. The purple points represent the value of the TD action QSL for our initial guess of constant $\dot{p}(t)$. The red points are the value of the TD action speed limit after optimising over all possible $\dot{p}(t)$ with the desired start and end points. These points lie on the blue line demonstrating that we can use optimal control to saturate equation~\eqref{CSineq}. \textbf{(b)} Shows the optimal function $p(t)$ for various values of $\theta$ as compared to our initial guess. \textbf{(c)} and \textbf{(d)} Are as for panels (a) and (b) except applied to the QFI speed limit. \textbf{(e)} and \textbf{(f)} Are as for panels (a) and (b) except applied to the WY speed limit.}
\label{fig:TD}
\end{figure}

In Fig.~\ref{fig:TD} we display the result of the numerical implementation of optimal control strategies on $\dot{p}(t)$ and their impact on the associated QSL. First, it is immediately evident from Fig.~\ref{fig:TD}(a) that different profiles of $p(t)$ (shown as dotted curves) lead to very different values of the action-QSL bound. In particular, since the path does not coincide with the geodesic, a constant ramp profile $p(t) = \frac{t}{\tau}$ (lower, purple dots) is clearly a non-optimal solution, as it results in a value markedly below the tightest theoretical bound given by the geometric QSL (blue solid curve), Eq.~\eqref{CSineq}. A fully optimized $\dot{p}(t)$ (red dots) demonstrates that we are able to saturate Eq.~\eqref{CSineq}. In contrast however, evaluating Eq.~\eqref{GEOMETRICQSL} using these two ramp profiles gives the same result (the solid blue curve in Fig.~\ref{fig:TD}) thus confirming that geometric QSLs are insensitive to the instantaneous speed. In Fig.~\ref{fig:TD}(b) we show the optimal profiles $p(t)$ which result in a constant metric speed, for different values of $\theta$, i.e. for different fixed paths. A completely analogous treatment can be implemented for any other metric, e.g. for the QFI and WY action introduced above. The optimal protocols for $p(t)$ will of course be different given choice and path as determined by $\theta$. This is shown in Fig.~\eqref{fig:TD}(c)-(f) for the QFI and for the WY metrics, thus demonstrating the general validity of our approach.
%that limit even when the selected dynamics is non-optimal (i.e. does not coincide in general with a geodesic path according to the TD metric, as is the case for all values of $\theta$ such that $\tau_{QSL}/\tau \neq 1$. This result represents a striking evidence of the possibility to exploit optimal control strategies in order to achieve the ultimate QSL.

%By applying the optimal control we should in theory be able to find a $\dot{p}(t)$ such that the Cauchy-Schwarz inequality in equation~\eqref{CSineq} is saturated. This is demonstrated in figure~\ref{fig:TD}(a) where we can see that our optimised action QSL is equal to the square of the geometric QSL. The initial, constant, guess for the optimal control parameter is significantly looser for most initial states. Some examples of the optimal control parameter $\dot{p}(t)$ are shown in figure~\ref{fig:TD}(b)

%In conclusion, the above analysis demonstrates in practice the power of our action-QSL, which allows not only to compare between different paths and different ways of travelling along it, but also to apply optimal control techniques in order to explicitly reach the QSL for the evolution time.

% \begin{figure}[t]
% (a) \hskip0.45\columnwidth(b)
% \includegraphics[width=0.45\columnwidth]{QFIactionPlot.pdf}~\includegraphics[width=0.45\columnwidth]{pQF.pdf}\newline
% (c) \hskip0.45\columnwidth(d)
% \includegraphics[width=0.45\columnwidth]{WYactPlot.pdf}~\includegraphics[width=0.45\columnwidth]{pWY.pdf}
% \caption{
% \textbf{(a)}The blue line shows the (square of the) geometric quantum Fisher information (QFI) speed limit for the generalised amplitude damping channel. This speed limit is independent of $p(t)$ as long as $\dot{p}(t)>0$. The purple points represent the value of the QFI action speed limit for our initial guess of constant $\dot{p}(t)$. The red points are the value of the QFI action speed limit after optimising over all possible $\dot{p}(t)$ with the desired start and end points. These points lie on the blue line demonstrating the we can use optimal control to saturate equation~\eqref{CSineq}. We choose $p(0) = 0.01$ and $p(1) = 0.99$ due to numerical infinities appearing at $p=0$ and $p=1$. \textbf{(b)} shows the optimal function $p(t)$ for various values of $\theta$ as compared to our initial guess.\textbf{(c)} and \textbf{(d)} show the exact same treatment for the WY speed limits.}
% \label{fig:QFI}
% \end{figure}

%%%%%%%%%%%%%%%%%%%%

%%%%%%%%%%%%%%%%%%%%%

\section{Conclusions}
\label{sec:Concl}
In this work we carefully assessed the geometric approach to QSLs in open quantum systems. If one has the freedom to choose any path connecting two states then any and all of the infinite families of geometric QSL are saturable. While the existence of a unique and tightest lower bound for the evolution time is guaranteed once a path and dynamics are completely specified, these constraints necessarily leave no room for optimization if the dynamics does not already coincide with a geodesic path. Therefore, care must be taken in interpreting open system geometric QSL times: they provide a quantitative indication of how far a given path is from the geodesic rather than necessarily indicating an achievable minimal time. We also highlighted that these geometric QSL times are insensitive to how this path is traversed and are therefore agnostic to the instantaneous speed. 

Nevertheless, this speed is an important and tunable degree of freedom. We have introduced a novel family of QSLs, termed action quantum speed limits, that explicitly depend on both the path and the instantaneous speed for a given metric. Our derivation relied on the same geometrical representation of quantum states and followed from the application of the Cauchy-Schwartz inequality to the path length. We established that the bound provided by the geometric QSL coincides with a special instance of the action QSL, specifically when the instantaneous speed of the latter is fully optimized along the path. We explicitly demonstrated this using optimal control techniques applied to a qubit undergoing a dynamics described in terms of a generalized amplitude damping channel for three choices of metric, the trace-distance, quantum Fisher information, and Wigner-Yanase skew information. While our formulation applies to arbitrary finite-dimensional systems, we expect that solving the optimal control problem becomes computationally more demanding for increasing system size. Our results provide a means to quantitatively assess the optimality of a given dynamical process from a purely geometric perspective. In addition, we have highlighted that the geometric formulation of quantum speed limits can be combined with optimal control techniques to characterise a particular dynamics; such an approach could be employed to find achievable minimal times for a given process~\cite{PoggiPRA,PoggiArXiv,DazV2020}. Our results may also be relevant to recent proposals employing optimal control in dynamical quantum estimation schemes~\cite{liu2017quantum,mirkin2020quantum,basilewitsch2020optimally,feyles2019dynamical}. Furthermore, our framework can naturally be extended to recently proposed resource speed limits~\cite{Campaioli2020, PiresArXiv}.

\acknowledgements
We are grateful to S. Deffner and B. Vacchini for useful discussions. E.O.C. and S.C. are supported by the Science Foundation Ireland Starting Investigator Research Grant ``SpeedDemon”, Grant No. 18/SIRG/5508. G.G acknowledges the European Research Council Starting Grant ODYSSEY Grant Agreement No. 758403 and FQXi and DFG Grant No. FOR2724.

\setcounter{equation}{0}
\renewcommand\theequation{A.\arabic{equation}}
\section*{Appendix A: Geodesic Paths}
The geodesic path for the QFI was derived by Uhlmann~\cite{Uhlmann1995} and is given by
\begin{equation}
\rho(t)=\frac{[(p(t)\omega_\tau+(1-p(t))\omega_0)((p(t)\omega_\tau^\dagger+(1-p(t))\omega^\dagger_0))]}{\| {p(t)\omega_\tau+(1-p(t))\omega_0} \|^2}
\end{equation}
where $\omega_0$ is a purification of $\rho(0)\!=\!\omega_0\omega_0^\dagger$. Therefore if $\rho(0)$ has a spectral decomposition $\rho(0) \!=\! \sum_i p_i \dyad{p_i}$ then we define $\omega_0 \! =\! \sum_i \sqrt{p_i} \ket{p_i}\bra{\phi_i}$ where ${\ket{\phi_i}}$ is another orthonormal basis of the Hilbert space, with $\omega_\tau$ defined in terms of $\omega_0$ as 
\begin{equation}
\omega_\tau = \rho(0)^{-1/2}(\rho(0)^{1/2}\rho(\tau)\rho(0)^{1/2})^{1/2}\rho(0)^{-1/2}\omega_0.
\end{equation}
Similarly for the WY metric the geodesic path was derived by Gibilisco~\cite{Gibilisco2003} and is given by
\begin{equation}
\rho(t)=\frac{\left((1-p(t))\sqrt{\rho(0)}+p(t)\sqrt{\rho(\tau)}\right)^2}{\tr{\left((1-p(t))\sqrt{\rho(0)}+p(t)\sqrt{\rho(\tau)}\right)^2}}.
\label{WYgeoeq}
\end{equation}
Finally for the TD, and in fact in any $p$-norm, the geodesic can be shown to be
\begin{equation}
\rho(t)=(1-p(t))\rho(0)+p(t)\rho(\tau).
\label{TRgeoeqAppendix}
\end{equation}
The proof of this is shown below. Deffner~\cite{Deffner2017} showed that for any Schatten-$p$-distance
\begin{equation}
    \mathcal{L}_p(\rho_0,\rho_\tau) = \left\lVert\,\rho_0 - \rho_\tau \right\rVert_p \equiv (\tr{|\,\rho_0 - \rho_\tau|^p})^{1/p}
\end{equation}
we have the inequality
\begin{equation}
     \dot{\mathcal{L}}_p(\rho_0,\rho_t) \leq \left\lVert\,\dot{\rho}_t \right\rVert_p.
\end{equation}
Integrating both sides of this inequality gives us a QSL of the form in Eq.~\eqref{path_len}. Substituting the geodesic path from Eq.~\eqref{TRgeoeqAppendix} into the right hand side of our inequality we arrive at
\begin{equation}
    \left\lVert\,\dot{p}(t)(\rho_\tau - \rho_0) \right\rVert_p = |\dot{p}_t| \mathcal{L}_p(\rho_0,\rho_\tau).
\end{equation}
Under the condition that $\dot{p}(t)\!\geq\! 0$ this path will saturate any of the infinite family of $p$-distance based QSLs. 

\section*{Appendix B: Optimal Control}
\renewcommand\theequation{B.\arabic{equation}}
Here we will give an overview of the optimal control techniques used in the calculation of the ramp profile $p(t)$ that minimises the action. For a more detailed explanation see, for example, Refs~\cite{glaser2015, Deffner2014}. We make use of Pontryagin's optimum principle~\cite{Kirk2004}. The cost functional that we want to minimise is the action along our path.
\begin{equation}
    a_{g}^{\gamma} \!=\! \int_0^\tau dt\,\sum_{jk=1}^M g_{jk}\frac{d\lambda_j}{dt}\frac{d\lambda_k}{dt} \equiv \int_0^\tau dt\,\mathscr{L}(\rho(t),\dot{\rho}(t),\dot{p}(t)).
\end{equation}
This equation defines the control Lagrangian, $\mathscr{L}(\rho(t),\dot{\rho}(t),\dot{p}(t))$ of the dynamics. $\dot{p}(t)$ is simply the time derivative of the ramp profile and can be thought of as the external control parameter. We then take the Legendre transform of the control Lagrangian to get the control Hamiltonian. This Hamiltonian is metric dependent and has no relation to the Hamiltonian that appears in the Schr\"odinger equation. We can now apply Pontryagin's optimum principle which states that control protocol $\dot{p}^*(t)$ that minimises our action is given by
\begin{equation}
    \mathcal{H}(\dot{p}^*(t)) = \underset{\dot{p}(t)\in \mathcal{A}}{\text{sup}}\mathcal{H}(\dot{p}(t)).
\end{equation}
Here $\mathcal{A}$ represents the set of all admissible profiles. In our case this corresponds to all profiles that map our pure initial state to the the thermal state i.e. $\int_0^\tau dt\, \dot{p}(t) = p(\tau) - p(0) = 1$. This allows us to find the optimal profile by a point-wise optimisation of the Hamiltonian rather than an optimisation over the full function space of $p(t)$. In order to perform the optimisation we use a modified gradient descent algorithm
\begin{equation}
    \dot{p}^{n+1}= \dot{p}^n + \epsilon\left(\frac{\partial \mathcal{H}^n}{\partial \dot{p}}-\frac{1}{\tau}\int_0^\tau dt\, \frac{\partial \mathcal{H}^n}{\partial \dot{p}}\right),
\end{equation}
and this method always results in an admissible profile. 

\bibliography{refs}

\end{document}